\documentclass[preprint2]{aastex}
\def\eg{{e.g.~}}

\shorttitle{AKARI Spectroscopy of BEGs}

\shortauthors{J.~H.~Lee et al.}

\begin{document}

\title{AKARI NEAR-INFRARED SPECTROSCOPY OF SDSS-SELECTED BLUE EARLY-TYPE GALAXIES}

\author{Joon Hyeop Lee $^{1}$, Ho Seong Hwang $^{2}$, Myung Gyoon Lee $^{3}$, Jong Chul Lee $^{3}$, and Hideo Matsuhara $^{4}$}
\affil{$^{1}$Korea Astronomy and Space Science Institute, Daejeon 305-348, Korea\\
$^{2}$CEA Saclay/Service d'Astrophysique, F-91191 Gif-sur-Yvette, France\\
$^{3}$Astronomy Program, Department of Physics and Astronomy, Seoul National University, Seoul 151-742, Korea\\
$^{4}$Institute for Space and Astronautical Science, Japan Aerospace and Exploration Agency, Sagamihara, Japan\\
}

\email{jhlee@astro.snu.ac.kr, hoseong.hwang@cea.fr, mglee@astro.snu.ac.kr, jclee@astro.snu.ac.kr, maruma@ir.isas.jaxa.jp
}

\begin{abstract}
A near-infrared (NIR; $2.5-4.5\mu$m) spectroscopic survey of SDSS(Sloan Digital Sky Survey)-selected blue early-type galaxies (BEGs) has been conducted using the AKARI. The NIR spectra of 36 BEGs are secured, which are well balanced in their star-formation(SF)/Seyfert/LINER type composition.
For high signal-to-noise ratio, we stack the BEG spectra all and in bins of several properties: color, specific star formation rate and optically-determined spectral type. We estimate the NIR continuum slope and the equivalent width of 3.29 $\mu$m PAH emission.
In the comparison between the estimated NIR spectral features of the BEGs and those of model galaxies, the BEGs seem to be old-SSP(Simple Stellar Population)-dominated metal-rich galaxies with moderate dust attenuation. The dust attenuation in the BEGs may originate from recent star formation or AGN activity and the BEGs have a clear feature of PAH emission, the evidence of current SF. BEGs show NIR features different from those of ULIRGs, from which we do not find any clear relationship between BEGs and ULIRGs.
We find that Seyfert BEGs have more active SF than LINER BEGs, in spite of the fact that Seyferts show stronger AGN activity than LINERs. One possible scenario satisfying both our results and the AGN feedback is that SF, Seyfert and LINER BEGs form an evolutionary sequence: SF $\rightarrow$ Seyfert $\rightarrow$ LINER.
\end{abstract}


\section{INTRODUCTION}

Early-type galaxies are often regarded as the objects at the final stage of galaxy evolution. Throughout many observations, it has been revealed that most nearby early-type galaxies have red colors and very poor contents of cold gas \citep[e.g.][]{fab76,tra00,tre06}. Such observational evidence indicates that early-type galaxies are composed of old stars and that they probably will not have additional star formation in the future. That is, most early-type galaxies seem to evolve \emph{passively}, unlike late-type galaxies with active star formation.

However, since \citet{abr99} and \citet{men99} found a considerable fraction of early-type galaxies with blue colors and evidence of current star-formation in the Hubble Deep Field \citep{wil96}, the existence of \emph{blue early-type galaxies} (hereafter, BEGs) has become a new important factor to be considered in the formation scenario of early-type galaxies. BEGs are known to have positive color gradients (i.e.~center bluer than outskirt) on average and consist of both old stars and young stars \citep{men01a,elm05,fer05,lee06,lee08,lee10a}. It has been also revealed that the internal color distributions of BEGs are neither homologous nor symmetric \citep{men04,lee06}, and that some BEGs are suspected to host active galactic nuclei (AGNs) \citep{men05,lee06,lee08}. Unlike typical red early-type galaxies (REGs) preferring high density environments, BEGs tend to reside in intermediate or low density environments \citep{kan09,lee10b}.
The origin and future of BEGs are still controversial. Observational evidence indicates that some BEGs may originate from recent galaxy interaction or mergers \citep{lee06}, but the gas infall is also a possible mechanism to build BEGs \citep{men01b}. After some time passes, BEGs may evolve into typical REGs \citep{lee06,lee07,lee08}, but it is also possible that they evolve into bulges of late-type galaxies \citep{ham05,kan09}.
In any case, BEGs are probably in the forming/growing phase of early-type galaxies or bulges.

To understand better the formation and evolution of BEGs, we need to inspect what happens in the inside of BEGs.
Since more than a half of BEGs have evidence of their merger/interaction origin \citep{lee06}, the interaction-induced star formation (SF) is expected to make BEGs blue. On the other hand, galaxy merger/interaction sometimes also causes AGN activity \citep{kew03,san05,com09}, and the power-law continua of AGNs may affect the color of BEGs.
\citet{lee08} reported that non-passive BEGs in the Sloan Digital Sky Survey \citep[SDSS;][]{yor00} include both SF galaxies and AGN host galaxies, the number ratio of which is about 3:5 for $-22.9<{^{0.1}M_{r}}$ \footnote{Absolute Petrosian magnitude in the $r$ band with K-correction as if the object were at z = 0.1 \citep{lee08}. All magnitudes in this paper are in the \emph{AB} system.} $<-20.7$.
However, optical spectra are easily obscured by dust, causing the missing of a significant fraction of AGNs in optical surveys \citep{bes05}.
Moreover, ground-based spectroscopy (\eg the SDSS spectroscopy) of bright galaxies in the optical band often suffers the dilution of low-contrast emission lines due to strong emission lines \citep{ho97}, which also causes inaccuracy in estimating AGN activity.

One of the most powerful tools to probe the SF/AGN activity is the spectroscopy in the infrared (IR) band.
Polycyclic aromatic hydrocarbons \citep[PAHs;][]{sel84,leg84} have been detected in various objects, such as H{\scriptsize II} region, protostellar clouds, planetary nebulae, and SF galaxies \citep{pee02,hon02}. It is known that UV photons excite the vibrational and stretching modes of PAH molecules, from the relaxation of which broad IR emission lines (such as 3.3, 6.2, 7.7, 8.6 and 11.2 $\mu$m lines) are produced \citep{che89,sha91}. Such PAH emission features are good indicators of SF activity.
On the other hand, \citet{voi92} argued that PAHs in the central region of an AGN can be destructed by the strong X-ray emission from the AGN. Furthermore, AGN activity is known to suppress SF itself \citep[AGN feedback;][]{ant08,raf08}.
Thus, those properties of PAH emission are believed to provide a reasonably clean diagnostic between SF and AGN.
In addition, the shape of the IR continuum is another indicator of SF/AGN, because SF galaxies and AGNs have different spectral energy distribution \citep[e.g.][]{ima08,ris06,ris10}.
However, no systematic IR spectroscopic surveys of BEGs are seen in the literature.

In this paper, we report the result of the first BEG spectroscopic survey in the near-IR (NIR) band, using the AKARI \citep{mur07} InfraRed Camera \citep[IRC;][]{ona07}.
The outline of this paper is as follows. Section 2 describes the target selection and the AKARI spectroscopic observation. Section 3 explains the reduction and analysis of the AKARI data. The extracted NIR spectra of BEGs and their features are shown in \S4.
The results are discussed in \S5, and the main results and their implication are summarized in \S6. Throughout this paper, we adopt
the cosmological parameters: $h=0.7$, $\Omega_{\Lambda}=0.7$, and
$\Omega_{M}=0.3$.

\section{OBSERVATION}

\subsection{Target Selection}

\begin{figure}[!t]
\plotone{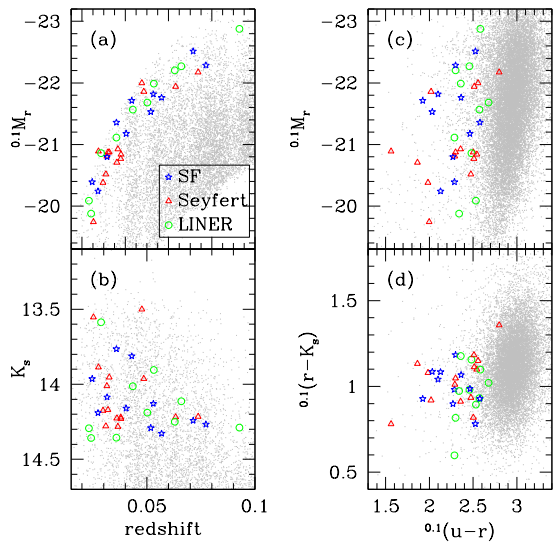}
\caption{ Basic optical-NIR properties of the observed 36 blue early-type galaxies (BEGs). (a) Absolute magnitude in the $r$ band as a function of redshift; (b) Apparent magnitude in the $K_s$ band as a function of redshift; (c) the color -- magnitude relation; and (d) the color -- color relation. The small dots are SDSS early-type galaxies and the large symbols are the BEGs: stars for star-forming (SF) BEGs, triangles for Seyfert BEGs and circles for LINER BEGs (classified in Fig.~\ref{bpt}).
\label{basic}}
\end{figure}

\begin{figure}[!t]
\plotone{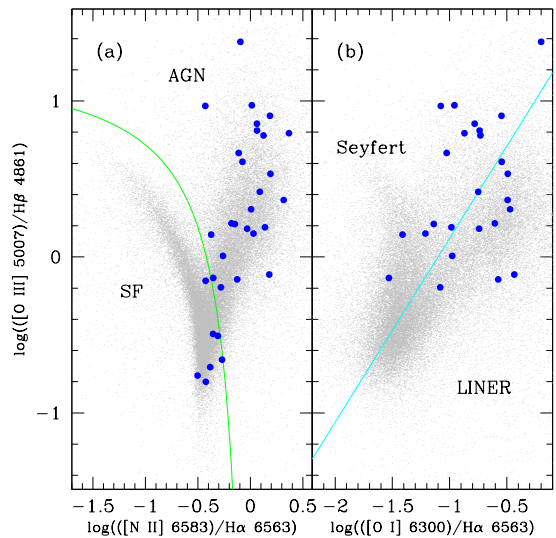}
\caption{ The distribution of SDSS early-type galaxies (small dots) and the BEGs (filled circle) in the line flux ratio diagram.
(a) The line is the boundary dividing SF galaxies (lower-left) and AGNs (upper-right), given by \citet{kau03}. (b) The line is the boundary dividing Seyferts (upper-left) and LINERs (lower-right), given by \citet{kew06}. The number of filled circles is smaller than our BEG sample size, because the BEGs with each line S/N $>3$ are plotted here. SF BEGs are not plotted in (b).
\label{bpt}}
\end{figure}

\begin{figure}[!t]
\plotone{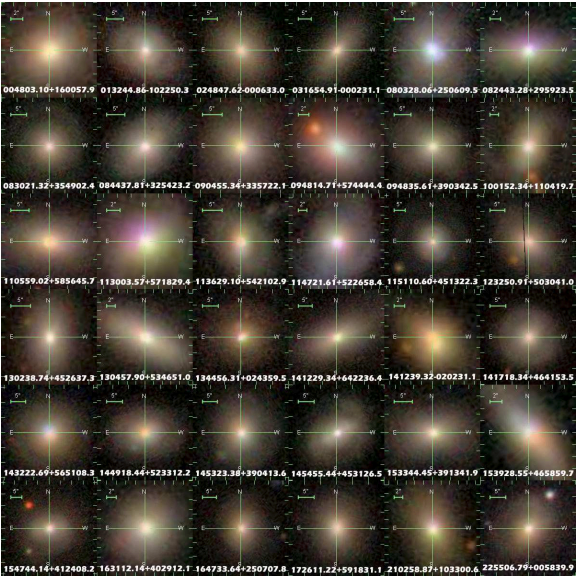}
\caption{ The atlas images of the observed 36 BEGs, retrieved from the SDSS. The bar in each image represents a scale of $2''$ or $5''$.
\label{atlas}}
\end{figure}

We selected 59 non-passive (i.e.~currently star-forming or hosting an AGN) BEG targets based on the scheme of \citet{lee08}, using the spectroscopic sample of galaxies in the SDSS Data Release 4 \citep[DR4;][]{ade06}.
We adopted the physical parameters of the galaxies from several value-added galaxy catalogs (VAGCs) drawn from the SDSS:
photometric parameters from the SDSS pipeline \citep{sto02}, structural parameters estimated by \citet{par05} and \citet{cho07}, and spectroscopic parameters from Max-Planck-Institute for Astrophysics (MPA) / Johns Hopkins University(JHU) VAGC \citep{kau03,tre04,gal06}.
Firstly, we selected the early-type galaxies from the SDSS galaxies using their distribution in the color, color gradient and light-concentration parameter space \citep{par05}.
Next, we divided the selected early-type galaxies into REGs and BEGs, using the method of \citet{lee06} based on the color distribution of bright early-type galaxies as a function of redshift \footnote{Here, BEGs are bluer than the Gaussian peak in the $(g-r)$ color distribution of the bright ($^{0.1}M_r<-20$) early-type galaxies, by more than 3$\sigma$, where $\sigma$ is the Gaussian spread.}.
The BEGs were classified using their flux ratios between several spectral lines \citep[e.g.~BPT diagram;][]{bal81,kew06} into passive, SF, Seyfert, and low ionization nuclear emission region (LINER) galaxies, among which passive BEGs were excluded.
Finally, we dropped BEGs with conspicuous sub-structures or disk components in visual checks.
Among the BEGs selected in this scheme, we chose the target objects as non-passive BEGs at $0.02<$~z~$<0.1$ with $K_{s}$ $<14.4$ \footnote{The `20 mag arcsec$^{-2}$ isophotal Ks fiducial ellipse aperture magnitude' from the Two Micron All Sky Survey \citep[2MASS;][]{jar00} Extended Source Catalog (XSC). http://www.ipac.caltech.edu/2mass/.}. The magnitude limit was applied to ensure the quality of the AKARI spectroscopy.

As a result, 59 BEGs were selected and proposed to be observed, but only 36 targets among them were successfully observed by the AKARI \footnote{ There is various unpredictability in the space telescope operation, which makes it difficult to carry out the observations exactly as planned. Because of such unpredictability, the perfect success of proposed observations is not guaranteed.}.
Fig.~\ref{basic} shows the basic optical-NIR properties of the 36 BEGs in our sample: their magnitudes as a function of redshift, color -- magnitude relation and color -- color relation, confirming that our sample BEGs are blue and relatively bright objects.
Fig.~\ref{bpt} displays the distribution of the BEGs in the line flux ratio diagrams, showing that the SF:Seyfert:LINER composition in our BEG sample is balanced well (11:15:10).
In Fig.~\ref{atlas}, the atlas images of the BEGs retrieved from the SDSS are presented.
The basic properties of the BEGs are summarized in Table~\ref{property}.

\subsection{AKARI Spectroscopy}

The AKARI \footnote{http://www.ir.isas.jaxa.jp/ASTRO-F/Outreach/index\_e.html} is a Japanese infrared astronomy satellite, from the Institute of Space and Astronautical Science (ISAS) of the Japanese Aerospace Exploration Agency (JAXA). It was launched on 21 February 2006 and ran out of its on-board coolant (Helium) supply on 26 August 2007 after its successful operation. The observation of our BEG targets was conducted during the AKARI open-time operation, the post-Helium phase of the AKARI observation, in which NIR ($1.8-5.5$ $\mu$m) imaging and spectroscopic observations are available.

The observation was conducted from November 2008 to August 2009.
The target BEGs were observed using the IRCZ4 Astronomical Observing Template (AOT), which is the spectroscopic AOT of the IRC. Between the grism and prism, we selected the grism spectroscopy that has better spectroscopic resolution.
The 1-$\sigma$ noise equivalent flux density of the IRCZ4 grism is about 0.2 mJy, and its spectroscopic resolution is R $\sim120$.
As the target positioning option, we used the point source aperture (Np), because the typical effective radius of target BEGs is only several arcseconds.
Among the successfully-observed 36 targets, 29 objects were observed twice, while seven targets only once. The net exposure time for a single pointing observation is about six minutes \citep{ona07,ima08}.

\section{DATA REDUCTION AND ANALYSIS}

\subsection{Basic Data Reduction and Median Stacking}

Unlike our expectation, two pointings per each object were not enough to secure sufficient signal-to-noise ratio (S/N) for most targets (a half of the spectra have $3.5-3.9\mu$m continua with S/N $<6$), although the spectra of a few objects are relatively good.
Thus, we applied the median stacking analysis technique for high S/N. The steps of the stacking analysis are as follows:

\begin{itemize}
\item[Step.1] We used the \emph{IRC Spectroscopy Toolkit for Phase 3 data Version 20090211} provided by the AKARI team, for the basic processes such as dark substraction, image flat fielding, spectral flat fielding, image combining, source detection and extraction, sky substraction, wavelength calibration and flux calibration.

\item[Step.2] Using the SDSS spectroscopic redshift, we converted the observer-frame wavelength into the rest-frame wavelength for each object. Since the rest-frame wavelength ranges are different between objects, we trimmed each spectrum with the rest-frame wavelength range of 2.5 -- 4.5 $\mu$m.

\item[Step.3] All trimmed spectra were normalized for the median flux density at $3.5 <\lambda< 3.9$ $\mu$m to be identical.

\item[Step.4] We stacked the normalized spectra by taking a median. We also produced several median-stacked spectra in bins of several properties such as optical spectral type, optical color, and [O{\scriptsize II}]-based specific star formation rate.
\end{itemize}

We selected the median stacking instead of mean stacking, because it removes unreal data points (hot or bad pixels) more effectively, whereas it improves data quality similarly to that the mean stacking does.
Not only the total stacking (i.e.~stacking all spectra into a single spectrum) but also subsample stacking was carried out to compare the NIR spectral features between subsamples with different optical properties.
We simply estimated the uncertainty of the stacked spectra, by supposing that the spectral noise follows the Poisson error and that the S/N is accumulated during the stacking process. That is,
\begin{equation}\label{errorstack}
(S/N)_s = \sqrt{\sum {(S/N)_i}^2},
\end{equation}
where (S/N)$_s$ is the S/N of the stacked spectrum and (S/N)$_i$ is the S/N of the individual spectra.
%{\it Caveats}. The uncertainty estimation through this way may not works well always. If we try to stack totally different spectra into a single spectrum, this uncertainty estimation returns meaningless values. In this paper, however, we stacked the spectra of objects classified into a single class with similar spectral energy distribution. Thus, the method described above can be used for a rough estimation of the spectral uncertainty, but it is needed to keep in mind that this is not a perfect uncertainty estimation.

\subsection{Spectral Fitting}

We used the power law continuum + 3.29 $\mu$m PAH emission (Gaussian) formula \citep{ris06} to fit the stacked spectra:
\begin{equation}\label{fitfn}
F_{\lambda} = A \lambda^{\Gamma} + B e^{-\frac{\scriptstyle{(\lambda-3.29 {\mu}m)^2}}{\scriptstyle{2{\sigma}^2}}} ,
\end{equation}
where $F_{\lambda}$ is the flux density, $\lambda$ is the wavelength in unit of $\mu$m, $\Gamma$ is the continuum slope, $\sigma$ is the Gaussian spread in unit of $\mu$m, and $A$ and $B$ are amplitudes. The fitting was conducted in the wavelength range of 2.5 -- 3.9 $\mu$m, because the Br$\alpha$ emission and CO absorption lines may affect the NIR spectra significantly at $\lambda >4\mu$m \citep{ima08,ris10}.

\section{RESULTS}\label{result}

\begin{figure}[!t]
\plotone{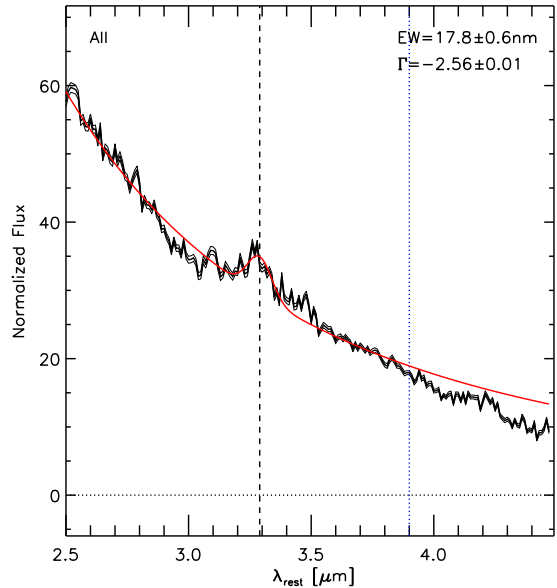}
\caption{ The stacked spectrum (noisy thick solid line) using all BEGs in our sample. The noisy thin solid lines show the S/N-accumulated noise and the smooth solid line is the continuum + PAH emission fit using Equation~(\ref{fitfn}). The estimated PAH equivalent width (EW$_{3.29}$) and the continuum slope ($\Gamma$) are denoted on the upper-right corner. The dashed vertical line shows the PAH central wavelength (3.29 $\mu$m), and the dotted vertical line shows the upper limit of the wavelength range for fitting (3.9 $\mu$m).
\label{specall}}
\end{figure}

Fig.~\ref{specall} displays the stacked spectrum of all BEGs in our sample. In this spectrum, a clear PAH emission line is found, the central wavelength of which is $3.29\mu$m. The continuum is overall fit well to the function in the fitting range ($2.5-3.9\mu$m). The ice-cored dust absorption ($\lambda\sim2.8-3.2\mu$m) feature is marginally found, but the bare carbonaceous dust absorption ($\lambda\sim3.4\mu$m) feature is hardly seen in Fig.~\ref{specall}. Around $\lambda\sim3.4\mu$m, instead of absorption features, a marginal excess is seen, which we currently have no idea of.
It is noted that the long-ward of 3.9 $\mu$m of the stacked spectrum shows values totally lower than the fit. The known line features at $\lambda>4\mu$m is Br$\alpha$ emission ($\lambda=4.05\mu$m) and CO$_2$ absorption ($\lambda=4.26\mu$m), but those features are not clearly identified in Fig.~\ref{specall}. The apparent `dip' at $\lambda>4\mu$m may be simply because the spectral fitting was carried out in the wavelength range of $2.5-3.9\mu$m.

\begin{figure}[!t]
\plotone{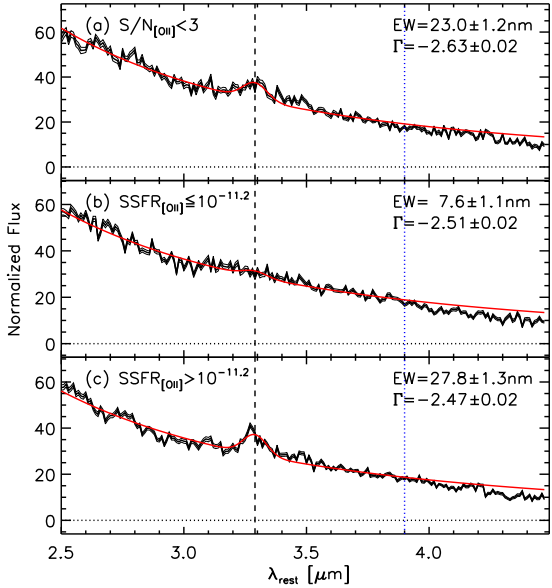}
\caption{ The stacked spectra of BEGs in three subsamples with different [O{\scriptsize II}]-based specific star formation rates (SSFRs). (a) S/N$_{[\textrm{\protect\tiny OII}]}<3$ (that is, SSFR$_{[\textrm{\protect\tiny OII}]}$ is not available); and among the BEGs with S/N$_{[\textrm{\protect\tiny OII}]}\ge3$: (b) SSFR$_{[\textrm{\protect\tiny OII}]}\le6.3$ and (c) SSFR$_{[\textrm{\protect\tiny OII}]}>6.3$, in unit of $10^{-12}$ yr$^{-1}$.
\label{specsfr}}
\end{figure}

\begin{figure}[!t]
\plotone{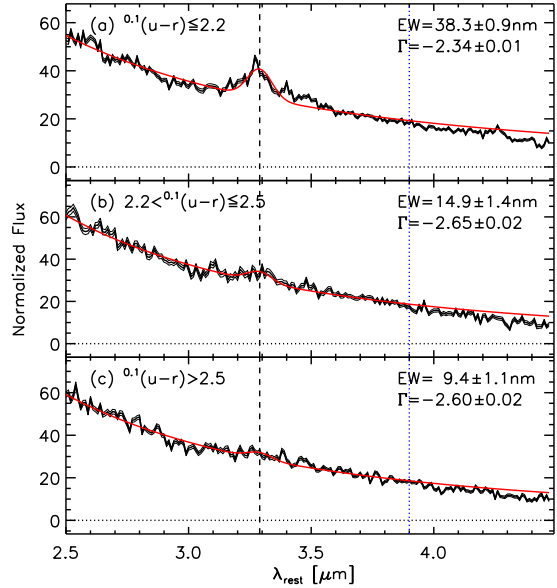}
\caption{ The stacked spectra of BEGs in three subsamples with different optical colors. (a) $^{0.1}(u-r)\le2.2$, (b) $2.2<{^{0.1}(u-r)}\le2.5$ and (c) $^{0.1}(u-r)>2.5$.
\label{specur}}
\end{figure}

\begin{figure}[!t]
\plotone{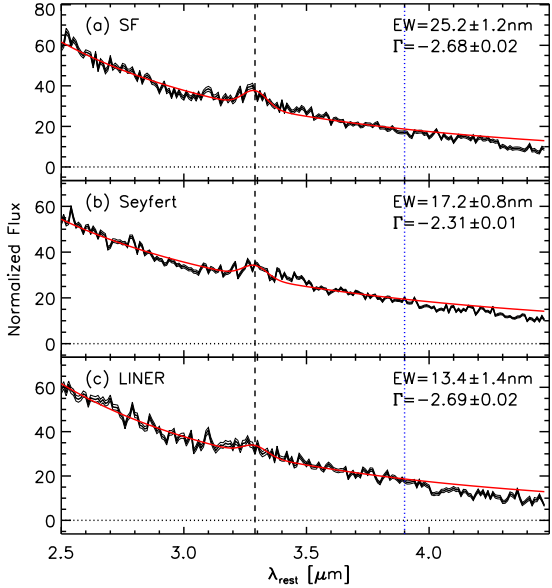}
\caption{ The stacked spectra of BEGs in three subsamples with different optical spectral types. (a) SF BEGs, (b) Seyfert BEGs and (c) LINER BEGs.
\label{spectype}}
\end{figure}

Figs.~\ref{specsfr} -- \ref{spectype} show the stacked spectra of the BEGs in bins of several properties: [O{\scriptsize II}]-based specific star formation rate (SSFR$_{[\textrm{\protect\tiny OII}]}$), $^{0.1}(u-r)$ color, and optical spectral type.
In the estimation of SSFR, the star formation rate (SFR) was estimated using the [O{\scriptsize II}] emission line that is known to be hardly contaminated by AGN emission \citep{ho05,kim06}, and the stellar mass was derived using the 2MASS $K_s$ magnitude \citep[see][]{lee10a}.
When considering the objects with sufficient [O{\scriptsize II}] emission signal (S/N$_{[\textrm{\protect\tiny OII}]}\ge3$; Fig.~\ref{specsfr}b and Fig.~\ref{specsfr}c), PAH emission is hardly found for low SSFR$_{[\textrm{\protect\tiny OII}]}$ BEGs, whereas the stacked spectum of the high SSFR$_{[\textrm{\protect\tiny OII}]}$ BEGs shows a clear PAH emission feature. This confirms the fact that PAH emission reflects current SF.
It is noted that 14 BEGs have low S/N$_{[\textrm{\protect\tiny OII}]}$, which may be mainly due to the difficulty in measuring the [O{\scriptsize II}](3727\AA) line using the SDSS spectroscopy ($3800-9200$\AA) at low redshift.

Fig.~\ref{specur} shows that the bluer BEGs have the stronger PAH emission feature, which indicates that the blue colors of BEGs reflect their SF activity. This trend is also confirmed in Fig.~\ref{spectype}: the PAH equivalent width (EW$_{3.29}$) of SF BEGs (25.2 nm) is larger than those of Seyfert and LINER BEGs (17.2 nm and 13.4 nm, respectively).
In Fig.~\ref{spectype}, it is noted that the EW$_{3.29}$ of Seyfert BEGs (17.2$\pm$0.8 nm) is larger than that of LINER BEGs (13.4$\pm$1.4 nm) by more than 2$\sigma$. Seyferts show stronger AGN activity than LINERs \citep[e.g.][]{gro06} and the current concensus on the relationship between AGN activity and PAH/SF is that AGN activity destroys PAH particles and suppresses SF \citep{voi92,ant08,raf08}.
Apparently, however, our results seem to be contradictory to such previous understanding (i.e.~Seyfert BEGs are more active than LINER BEGs in both AGN and SF activity).

\section{DISCUSSION}

\subsection{The Identity of BEGs from the AKARI View}

%\begin{figure}
%\plotone{f8.eps}
%\caption{ Synthesized spectra using the simple stellar population model (with solar metallity) of \citet{bru03}: (a) 0.1 Gyr old, (b) 3 Gyr old and (c) 10 Gyr old. The different line types indicate different dust attenuation levels: $\tau_V$ (the optical depth in the $V$ band) $=0$ (solid), $\tau_V=1$ (long-dashed), $\tau_V=5$ (short-dashed) and $\tau_V=10$ (dotted). Dust attenuation was estimated using the simple model of \citet{cha00}.
%\label{model}}
%\end{figure}

\begin{figure}[!t]
\plotone{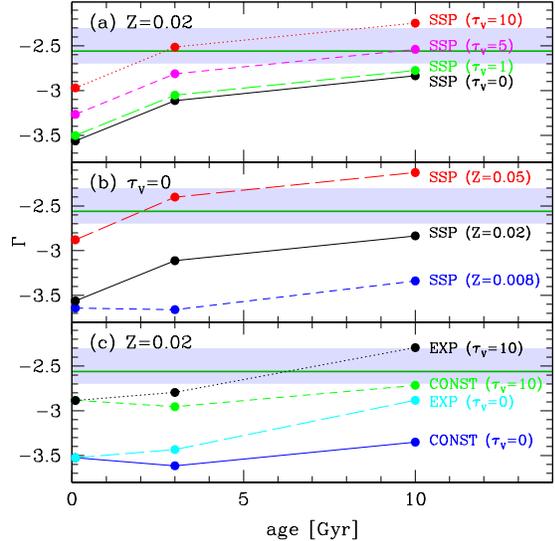}
\caption{ Comparison of NIR continuum slope ($\Gamma$) between population synthesis models and the stacked spectra of BEGs: (a) the simple stellar population (SSP) models with solar metallicity (Z=0.02) and various dust attenuation values (solid, long-dashed, short-dashed and dotted lines for $\tau_V$ $=0,1,5$ and 10, where $\tau_V$ is the dust optical depth in the $V$ band); (b) the SSP models with $\tau_V=0$ and various metallicity (short-dashed, solid and long-dashed lines for Z=0.008, 0.02 and 0.05; and (c) the exponentially-decreasing star formation rate model with the exponential time scale of 1 Gyr (EXP; long-dashed and dotted lines for $\tau_V=0$ and 10) and the model with constant star formation rate of 1 M$_{\odot}$/yr (CONST; solid and short-dashed lines for $\tau_V=0$ and 10), with Z=0.02. The shaded area shows the $\Gamma$ range of the BEG stacked spectra and the horizontal solid line shows the $\Gamma$ value of the all-BEG-stacked spectrum. Dust attenuation was estimated using the simple model of \citet{cha00}.
\label{modcomp}}
\end{figure}

%\begin{deluxetable}{lccc}
%\tablenum{2} \tablecolumns{4} \tablecaption{ Model spectra slopes \label{modelslope}} \tablewidth{0pt}
%\tablehead{ Model  & 0.1 Gyr & 3 Gyr & 10 Gyr }
%\startdata
%SSP$^{(1)}$ ($\tau_V=0$) & $-3.56$ & $-3.11$ & $-2.83$ \\
%SSP ($\tau_V=1$) & $-3.50$ & $-3.05$ & $-2.51$ \\
%SSP ($\tau_V=10$) & $-2.97$ & $-2.51$ & $-2.25$ \\
%EXP$^{(2)}$ & $-3.53$ & $-3.43$ & $-2.88$ \\
%CONST$^{(3)}$ & $-3.52$ & $-3.62$ & $-3.35$ \\
%\enddata
%\tablecomments{ (1) The simple stellar population model. (2) The exponentially-decreasing star formation rate model with the exponential time-scale of $\tau=1$ Gyr. (3) The constant star formation rate (1 M$_{\odot}/yr$) model. }
%\end{deluxetable}

BEGs are known to have not only very young stars but also a considerable amount of old stars \citep{men01a,fer05,lee10a}.
In our results, however, it is not easy to check this previous knowledge independently, because the NIR continuum slope ($\Gamma$) depends on both stellar and dust contents (dust either in host galaxies or in AGN tori).
Fig.~\ref{modcomp} shows the variation of $\Gamma$ as a function of age, metallicity, star formation history and dust attenuation, using the models of \citet{bru03}: simple stellar population (SSP), exponentially-decreasing and constant star formation rate models.
In Fig.~\ref{modcomp}, it is found that $\Gamma$ becomes less negative as star formation ended earlier, as age increases, as metallicity increases and as dust increases.
Due to this degeneracy, it is difficult to tell the dominant factor in determining $\Gamma$ of BEGs. Even if we suppose that the NIR continuum of the BEGs is old-SSP-dominated from the previous knowledge \citep[the age of most stars in BEGs is about 10 Gyr; e.g.][]{fer05,lee10a}, the $\Gamma$ of the BEGs can be reproduced using either the metal-rich SSP model without dust or the significantly dust-attenuated SSP model with solar metallicity.

\begin{figure}[!t]
\plotone{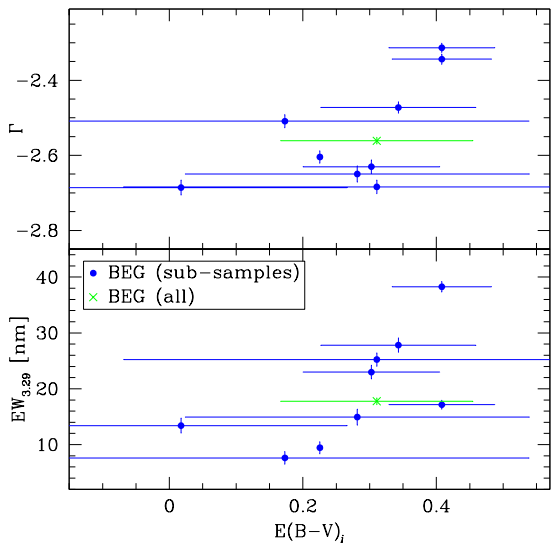}
\caption{ The internal extinction E($B-V$)$_i$ versus the $\Gamma$ and EW$_{3.29}$ of the BEGs. Filled circles are the values from the subsample-stacked spectra and cross is the value from the all-BEGs-stacked spectrum. The horizontal errorbars indicate the sample inter-quartile range for the median E($B-V$)$_i$ value in each subsample.
\label{balmer}}
\end{figure}

The effect of internal dust extinction is independently checked using the information extracted from the SDSS data; that is, using the Balmer decrement. Based on the formula of \citet{cal94}, we derived the internal reddening E($B-V$)$_i$ (median value for each subsample) and compared them with the $\Gamma$ and EW$_{3.29}$ of the stacked BEGs in Fig.~\ref{balmer}.
It is found that the NIR spectral parameters have correlations with E($B-V$)$_i$, in the sense that the $\Gamma$ and EW$_{3.29}$ increase with increasing E($B-V$)$_i$.
The E($B-V$)$_i$ values of the BEGs are not small, but seem to be not large enough to explain the $\Gamma$ of the BEGs in Fig.~\ref{modcomp} \citep[the $\tau_V$ corresponding to E($B-V$) = 0.4 is 1.14;][]{ruy05}. Thus, if we suppose moderate (not too strong) dust attenuation in the BEGs, Fig.~\ref{modcomp} indicates that the BEGs may have stellar populations with slightly high metallicity, although this estimation is strongly model-dependent.

\begin{figure}[!t]
\plotone{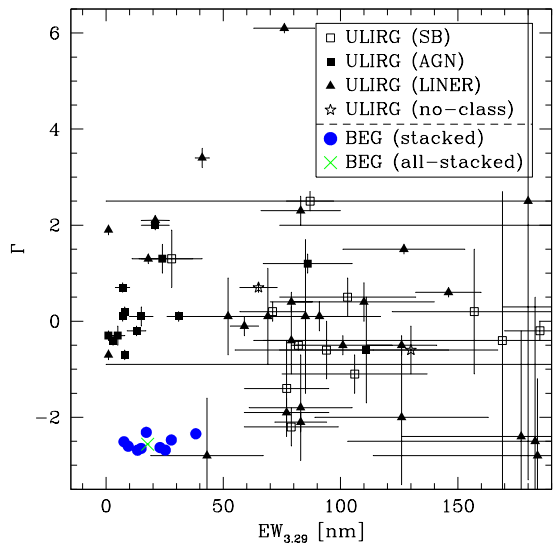}
\caption{ Comparison between BEGs and other objects in the $\Gamma$ -- EW$_{3.29}$ plane. The data for starburst ULIRGs (SB ULIRGs; open rectangles), AGN ULIRGs (filled rectangles), LINER ULIRGs (filled triangles) and not-classified ULIRGs (open stars)  are from \citet{ris10}. The filled circles are BEGs stacked in bins of several properties and the cross shows the all-BEGs-stacked result.
\label{risal}}
\end{figure}

Meanwhile, BEGs are often thought to be the intermediate objects transforming from interacting galaxies or mergers to red elliptical galaxies or bulges of late-type galaxies \citep{lee06,lee08,kan09}. Known to have similar origins are Ultra-Luminous InfraRed Galaxies (ULIRGs). ULIRGs are very luminous in the IR band due to vigorous starburst and/or AGN activity \citep{vei95,vei99a}, and are thought to be formed by strong interaction or merger of two disk galaxies \citep{san88}.
\citet{ris06,ris10} estimated the $\Gamma$ and EW$_{3.29}$ of some ULIRGs, showing an interesting division between AGN-host ULIRGs and starburst ULIRGs.

Fig.~\ref{risal} shows the loci of our BEGs on the $\Gamma$ -- EW$_{3.29}$ plane, compared with the objects presented in \citet{ris10}.
In this figure, it is found that our BEGs have $\Gamma$ and EW$_{3.29}$ values distinct from those of the ULIRGs, in the sense that the BEGs have largely negative $\Gamma$ values ($\sim-2.5$), while those of the ULIRGs are mostly larger than $-2$ (up to $6$).
The EW$_{3.29}$ of the BEGs is smaller than that of the ULIRGs on average, showing that the SF in the BEGs is not as active as that in the ULIRGs.
The $\Gamma$ value is expected to be largely positive when the NIR continuum is dominated by dust emission. In our BEG sample, however, even the stacked spectra using optical AGN or optical SF BEGs has largely negative $\Gamma$. This indicates that the dust amount in those BEGs may not be as large as that in ULIRGs, as shown already using the internal dust reddening derived from Balmer decrement \citep[the mean E$(B-V)$ value of ULIRGs is larger than 1.0;][]{vei99a,vei99b}.

In short, the $\Gamma$ and EW$_{3.29}$ features of the BEGs are not easy to interpret, because they are affected by the combined effects of age, metallicity, star formation history and dust attenuation. However, based on some previous knowledge, Balmer line information and a few assumptions, the BEGs are thought to be old-SSP-dominated metal-rich galaxies with moderate dust attenuation. The dust attenuation in the BEGs may originate from recent star formation or AGN activity. This interpretation is consistent with the previous understanding of stellar populations in BEGs \citep[that is, mostly old stars + partially young stars;][]{fer05,lee10a}.
There is a possibility that ULIRGs are the progenitors of BEGs in the merging/interacting phase \citep{vei02}, but
any clear evidence of the close relationship between BEGs and ULIRGs is not yet found in our results based on the NIR spectroscopy.

\subsection{NIR Features and AGN Activity in BEGs}

\begin{figure}[!t]
\plotone{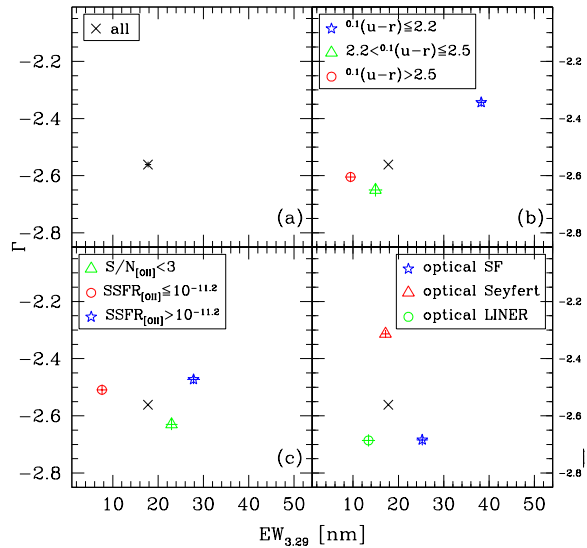}
\caption{ The $\Gamma$ versus EW$_{3.29}$ of the BEGs stacked in bins of several properties: (a) all, (b) $^{0.1}(u-r)$ color, (c) SSFR$_{[\textrm{\protect\tiny OII}]}$ and (d) optical spectral type. In every panel, the all-BEGs-stacked result is denoted as a cross.
\label{risal2}}
\end{figure}

As introduced in \S1, NIR spectroscopy provides interesting clues to the internal evolutionary process of the BEGs, related to their SF or AGN activity. Those clues become more useful when combined with the optical properties of the BEGs.
Fig.~\ref{risal2} compares the stacked spectra in bins of several properties on the $\Gamma$ -- EW$_{3.29}$ plane, giving a summary of the key results in Fig.~\ref{specall} -- \ref{spectype}.
The EW$_{3.29}$ reflects the current SF of the BEGs well (Fig.~\ref{risal2}c), but it is noted that the Seyfert BEGs have EW$_{3.29}$ larger than that of the LINER BEGs. As mentioned at the end of \S\ref{result}, this result is not well explained simply by the current concensus that AGN activity destroys PAH particles.

In addition to EW$_{3.29}$, another key parameter is $\Gamma$.
As shown in Fig.~\ref{modcomp}, $\Gamma$ is affected commonly by age, metallicity and dust.
However, the bluest BEGs with the largest (least negative) $\Gamma$ in Fig.~\ref{risal2}b is not easily understood by the age or metallicity effects, because blue galaxies tend to be young or metal-poor, which will make $\Gamma$ more negative unlike the bluest BEGs. Thus, the $\Gamma$ difference between the bluest BEGs and the other BEGs seems to be mainly due to the difference in their dust contents.
It is also noted that the optical SF BEGs have small (largely negative) $\Gamma$ in Fig.~\ref{risal2}d, which shows good consistency with the star formation history dependence of $\Gamma$, indicating that those SF BEGs are not very dusty.
That is, several AGNs with dusty tori in the bluest BEG subsample seem to mainly contribute to the largest (least negative) $\Gamma$.

In Fig.~\ref{risal2}c, the BEGs with large SSFR$_{[\textrm{\protect\tiny OII}]}$ have $\Gamma$ intermediate between those of the bluest BEGs and optical SF BEGs. These are because the optical color, SSFR and optical spectral type of the BEGs do not tightly correlate: some non-SF BEGs have large SSFR$_{[\textrm{\protect\tiny OII}]}$ or blue $^{0.1}(u-r)$ color.
These results lead us again to the conclusion that the SF and the AGN activity in BEGs are not necessarily contradictory to each other. The coexistence of SF and AGN is not entirely new discovery \citep[e.g.][]{tza07,tre09}, but currently it is widely believed that AGN activity tends to destroy PAH particles and to truncate SF \citep{voi92,wei06,sch07,tor09}. In our result, however, Seyfert BEGs have larger EW$_{3.29}$ than LINER BEGs, in spite of the fact that Seyferts show stronger AGN activity than LINERs.

This result may be explained by various ways.
\citet{smi07} and \citet{ima08} showed that PAHs can survive even if they are close to AGNs when the AGNs are not dust-free, which gives an answer to how the AGN activity and PAH emission coexist. However, it does not sufficiently explain why stronger AGNs have stronger PAH emission features.
One possibility is that Seyfert BEGs and LINER BEGs may not be the separate branches in BEG evolution. Satisfying our results, one possible scenario of BEG evolution is as follows \citep[see also][]{lee10a}:
\begin{itemize}
\item[Ph.1] SF BEG phase. Triggered by some mechanisms \citep[possibly interaction or merger;][]{lee06}, active (but not very dusty) SF occurs in the progenitor of a BEG, originally consisting of old stars in the main. The NIR continuum is very similar to that of old passive galaxies, but the PAH emission is clearly detected.
\item[Ph.2] Seyfert BEG phase. After the central black hole in a BEG increases its mass by gas accretion, AGN activity starts. The SF in the BEG starts to be suppressed by the AGN feedback, but at this time, both AGN and SF activities coexist. The NIR continuum is contaminated by the AGN ($\Gamma$ increases), and the PAH emission becomes weaker than that of SF BEGs.
\item[Ph.3] LINER BEG phase. The AGN feedback has truncated most SF by removing ambient gas in the BEG. Since this makes the gas accretion into the central black hole stop also, AGN activity becomes weak as a natural result. The NIR continuum is intermediate between SF BEGs and Seyfert BEGs, and the PAH emission feature is weakest among the three phases.
\end{itemize}
After these three phases, the BEGs may evolve into REGs or bulges of late-type galaxies via the phase of passive BEGs \citep{lee06,lee08,kan09}.
The proposed scenario is consistent with the findings about the relationship between Seyferts and LINERs in several recent studies \citep{sch07,sch09,hic09,lee10a}.

\section{SUMMARY}

We conducted an AKARI NIR spectroscopic survey of SDSS-selected BEGs. We secured the NIR spectra of 36 BEGs that are well balanced in their SF/Seyfert/LINER composition.
We stacked the BEG spectra all and in bins of several properties, such as color, SSFR and optical spectral type. This is the first presentation of the BEG NIR (2.5 -- 4.5 $\mu$m) spectra, from which we estimated the NIR continuum slope ($\Gamma$) and the equivalent width of 3.29 $\mu$m PAH emission (EW$_{3.29}$).

In the comparison between the estimated NIR spectral features of the BEGs and those of the model galaxies, the BEGs seem to be old-SSP-dominated metal-rich galaxies with moderate dust attenuation. The dust attenuation in the BEGs may originate from recent star formation or AGN activity and the BEGs have a clear feature of PAH emission, the evidence of current SF.
BEGs show NIR features different from those of ULIRGs, from which we can not find any clear relationship between BEGs and ULIRGs.

We compared the stacked spectra of the BEGs in bins of several properties, confirming that the PAH emission is a good indicator of current SF. We found that Seyfert BEGs have more active SF than LINER BEGs, in spite of the fact that the AGN activity in Seyfert BEGs are stronger than that in LINER BEGs.
One possible scenario satisfying both this result and the AGN feedback is that SF, Seyfert and LINER BEGs form an evolutionary sequence: SF $\rightarrow$ Seyfert $\rightarrow$ LINER.

\acknowledgments

JHL is a member of the Dedicated Researchers for Extragalactic AstronoMy (DREAM) in Korea Astronomy and Space Science Institute (KASI).
This paper is based on the data produced in the AKARI open-time phase-3 project, \emph{Infrared Spectroscopy of Blue Early-type Galaxies} (ISBEG; PI: JHL).
The AKARI is a JAXA project with the participation of ESA. The authors appreciate the help and support of the all AKARI project members.
The authors are grateful to an anonymous referee for useful comments.
This work was supported by the National Research Foundation of Korea (NRF) grant funded by the Korea Government (MEST) (No.
R01-2007-000-20336-0).
Funding for the SDSS and SDSS-II has been provided by the Alfred P. Sloan Foundation, the Participating Institutions, the National Science Foundation, the US Department of Energy, the National Aeronautics and Space Administration, the Japanese Monbukagakusho, the Max Planck Society, and the Higher Education Funding Council for England. The SDSS Web site is http://www.sdss.org/. The SDSS is managed by the Astrophysical Research Consortium for the Participating Institutions. 
The Participating Institutions are the American Museum of Natural History, Astrophysical Institute Potsdam, the University of Basel, the University of Cambridge, Case Western Reserve University, the University of Chicago, Drexel University, Fermilab, the Institute for Advanced Study, the Japan Participation Group, Johns Hopkins University, the Joint Institute for Nuclear Astrophysics, the Kavli Institute for Particle Astrophysics and Cosmology, the Korean Scientist Group, the Chinese Acadeour of Sciences (LAMOST), Los Alamos National Laboratory, the Max-Planck-Institute for Astronomy (MPIA), the Max Planck Institute for Astrophysics (MPA), New Mexico State University, Ohio State University, the University of Pittsburgh, the University of Portsmouth, Princeton University, the US Naval Observatory, and the University of Washington.
This publication makes use of data products from the 2MASS, which is a joint project of the University of Massachusetts and the Infrared Processing and Analysis Center/California Institute of Technology, funded by the National Aeronautics and Space Administration and the National Science Foundation.

\begin{deluxetable}{ccccccr@{.}l}
\tablenum{1} \tablecolumns{6} \tablecaption{Basic Properties of
the BEGs \label{property}} \tablewidth{0pt}

\tablehead{SDSS ID  & Redshift & $^{0.1}M_r$ & $K_s$ & Optical  & $^{0.1}(u-r)$ & \multicolumn{2}{c}{SSFR$_{[\textrm{\protect\tiny OII}]}$} \\
 &  & (1) & (2) & Spectral Type &  & \multicolumn{2}{c}{(3)} }

\startdata
004803.10+160057.9 & 0.050 & $-21.681$ & 14.189 & LINER & 2.676 & 0&000 \\
013244.86$-$102250.3 & 0.032 & $-20.876$ & 14.171 & Seyfert & 2.503 & \multicolumn{2}{c}{---} \\
024847.62$-$000633.0 & 0.025 & $-20.392$ & 13.964 & SF & 2.287 & \multicolumn{2}{c}{---} \\
031654.91$-$000231.1 & 0.023 & $-20.087$ & 14.293 & LINER & 2.532 & \multicolumn{2}{c}{---} \\
080328.06+250609.5 & 0.028 & $-20.888$ & 13.886 & Seyfert & 1.563 & \multicolumn{2}{c}{---} \\
082443.28+295923.5 & 0.025 & $-19.743$ & 13.553 & Seyfert & 1.998 & \multicolumn{2}{c}{---} \\
\hline
083021.32+354902.4 & 0.053 & $-21.818$ & 14.129 & SF & 2.099 & 11&989 \\
084437.81+325423.2 & 0.032 & $-20.808$ & 14.011 & Seyfert & 2.290 & 25&838 \\
090455.34+335722.1 & 0.044 & $-21.566$ & 14.013 & LINER & 2.572 & 0&306 \\
094814.71+574444.4 & 0.030 & $-20.382$ & 14.175 & Seyfert & 1.984 & \multicolumn{2}{c}{---} \\
094835.61+390342.5 & 0.071 & $-22.515$ & 14.242 & SF & 2.526 & \multicolumn{2}{c}{---} \\
100152.34+110419.7 & 0.052 & $-21.531$ & 14.291 & SF & 2.033 & 14&902 \\
\hline
110559.02+585645.7 & 0.048 & $-21.997$ & 13.501 & Seyfert & 2.557 & 1&804 \\
113003.57+571829.4 & 0.036 & $-20.708$ & 14.227 & Seyfert & 1.863 & 37&973 \\
113629.10+542102.9 & 0.063 & $-21.939$ & 14.218 & Seyfert & 2.513 & 9&908 \\
114721.61+522658.4 & 0.049 & $-21.858$ & 13.964 & Seyfert & 2.019 & 55&042 \\
115110.60+451322.3 & 0.036 & $-21.113$ & 14.356 & LINER & 2.287 & 2&687 \\
123250.91+503041.0 & 0.063 & $-22.204$ & 14.250 & LINER & 2.299 & 6&698 \\
\hline
130238.74+452637.3 & 0.024 & $-19.878$ & 14.358 & LINER & 2.340 & \multicolumn{2}{c}{---} \\
130457.90+534651.0 & 0.029 & $-20.859$ & 13.587 & LINER & 2.481 & 0&014 \\
134456.31+024359.5 & 0.077 & $-22.287$ & 14.267 & SF & 2.299 & \multicolumn{2}{c}{---} \\
141229.34+642236.4 & 0.036 & $-21.358$ & 13.765 & SF & 2.578 & \multicolumn{2}{c}{---} \\
141239.32$-$020231.1 & 0.074 & $-22.175$ & 14.215 & Seyfert & 2.798 & \multicolumn{2}{c}{---} \\
141718.34+464153.5 & 0.038 & $-20.839$ & 14.229 & Seyfert & 2.542 & 1&952 \\
\hline
143222.69+565108.3 & 0.043 & $-21.713$ & 13.813 & SF & 1.925 & 9&136 \\
144918.44+523312.2 & 0.066 & $-22.270$ & 14.113 & LINER & 2.457 & 3&405 \\
145323.38+390413.6 & 0.032 & $-20.798$ & 14.086 & SF & 2.270 & \multicolumn{2}{c}{---} \\
145455.44+453126.5 & 0.037 & $-20.924$ & 14.283 & Seyfert & 2.356 & 0&000 \\
153344.45+391341.9 & 0.040 & $-21.176$ & 14.160 & SF & 2.463 & 0&365 \\
153928.55+465859.7 & 0.038 & $-20.770$ & 14.220 & Seyfert & 2.509 & \multicolumn{2}{c}{---} \\
\hline
154744.14+412408.2 & 0.033 & $-20.873$ & 13.954 & Seyfert & 2.299 & 35&112 \\
163112.14+402912.1 & 0.031 & $-20.516$ & 14.279 & Seyfert & 2.476 & \multicolumn{2}{c}{---} \\
164733.64+250707.8 & 0.057 & $-21.762$ & 14.328 & SF & 2.363 & 6&926 \\
172611.22+591831.1 & 0.027 & $-20.241$ & 14.191 & SF & 2.128 & \multicolumn{2}{c}{---} \\
210258.87+103300.6 & 0.093 & $-22.877$ & 14.289 & LINER & 2.583 & 0&193 \\
225506.79+005839.9 & 0.053 & $-21.988$ & 13.905 & LINER & 2.359 & 0&046 \\
\enddata
\tablecomments{ (1) The $r$-band absolute magnitude with K-correction (Petrosian magnitude), (2) the $K_s$-band apparent magnitude (20 mag arcsec$^{-2}$ isophotal $K_s$ fiducial ellipse aperture magnitude), and (3) the specific star formation rate in unit of $10^{-12}$ yr$^{-1}$ (that is, the newly-forming stellar mass per unit stellar mass per year). For the objects with S/N$_{[\textrm{\protect\tiny OII}]}<3$, their SSFRs were not estimated (---). }
\end{deluxetable}

\end{document}